\documentclass{article}

\usepackage[preprint]{neurips_2024}

\usepackage[utf8]{inputenc} % allow utf-8 input
\usepackage[T1]{fontenc}    % use 8-bit T1 fonts
\usepackage{hyperref}       % hyperlinks
\usepackage{url}            % simple URL typesetting
\usepackage{booktabs}       % professional-quality tables
\usepackage{amsfonts}       % blackboard math symbols
\usepackage{nicefrac}       % compact symbols for 1/2, etc.
\usepackage{microtype}      % microtypography
\usepackage{xcolor}         % colors
\usepackage{graphicx}
\usepackage{amsmath}
\usepackage{subcaption}
\usepackage{multirow}

\title{A physics-enhanced multi-modal fused neural network for predicting contamination length interval in pipeline}

\author{Jian Du$^1$\thanks{jiandu1997@163.com}, Pengtao Niu$^1$, Jianqin Zheng$^2$, Qi Liao$^1$\thanks{qliao@cup.edu.cn}, Yongtu Liang$^1$\\
  $^1$China University of Petroleum-Beijing\\
  $^1$Fuxue Road No. 18, Changing District, Beijing 102249, PR China \\
  $^2$China Petroleum Planning and Engineering Institute \\
  $^2$Xinxi Road No 3, Haidian District, Beijing 100083, PR China \\
}

\begin{document}

\maketitle

\begin{abstract}
During the operation of a multi-product pipeline, an accurate and effective prediction of contamination length interval is the central key to guiding the cutting plan formulation and improving the economic effect. However, the existing methods focus on extracting implicit principles and insufficient feature correlations in a data-driven pattern but overlook the potential knowledge in the scientific theory of contamination development, may cause practically useless results. Consequently, in this study, the holistic feature correlations and physical knowledge are extracted and integrated into the neural network to propose a physics-enhanced adaptive multi-modal fused neural network (PE-AMFNN) for contamination length interval prediction. In PE-AMFNN, a multi-modal adaptive feature fusion module is created to establish a comprehensive feature space with quantified feature importance, thus capturing sufficient feature correlations. Subsequently, a mechanism-coupled customized neural network is designed to incorporate the explicit scientific principle into the forward and backward propagation. Besides, a physics-embedded loss function, which introduces interval differences and interval correlation constraints, is established to unearth the latent physical knowledge in contamination development and force the model to draw physically unreasonable results. Validation on the real-world cases implies that the proposed model outperforms the start-of-art techniques and latest achievements, with Root Mean Squared Relative Errors reduced by 31\% and 36\% in lower and upper limit prediction. Furthermore, the sensitivity analysis of model modules suggests that both the multi-modal feature fusion and the physical principle are crucial for model improvements.
\end{abstract}

\section{Introduction}

\subsection{Background}

As an inseparable and irreplaceable energy resource, refined oil has been widely applied in metallurgy, chemical industry, transportation, agricultural production, and other fields (\cite{DU2023127452}). To ensure a huge amount of refined oil can be transported to the markets, pipelines are becoming a primary means of refined oil transportation with safety and efficiency (\cite{WANG2020121831}) due to their exclusive mode of delivering different products continuously (\cite{WANG2021468}). Recently, the downstream market present a demand pattern of multiple batches and small volumes (\cite{BAMOUMEN2023106082}). As a result, sequential transportation is adopted given the economy and efficiency (\cite{LU2023105073}), which means different kinds of refined oil are delivered by sharing one pipeline (\cite{ZHENG2021510}).

When transporting different refined products back-to-back in one pipeline, a mixing interface between two adjacent oil batches continuously forms unavoidably (\cite{YUAN2023100105}). The formation of contamination production causes the degradation of product quality, which in turn leads to economic loss and quality accidents (\cite{YU2022107613}). To treat the contamination product, the delivery stations make a cutting plan according to the prediction of the contamination length (\cite{ABDELLAOUI2021107483}). However, there exists fluctuation in the quality of oil products from the refinery, leading to unqualified oil products in the mixed oil trailing. When the quality of the refined products from the refinery decreases, to prevent quality deterioration which is caused by mixed oil trailing, the station operators delay cutting time by three or five minutes when the mixed oil concentration tends to be stable (\cite{doi:10.1177/0144598720911158}). This results in an issue that the contamination length exhibits intensive fluctuation characteristics (\cite{LU2023162386}), which brings trouble for station operators to make a reasonable receiving plan for contamination only relying on the specific prediction of contamination length (\cite{pr7010007}). Therefore, it is urgently needed to provide an effective prediction tool for contamination length fluctuation intervals.

The variation of contamination length fluctuation interval is highly related to the formation process of contamination which is essentially a complex mechanism process (\cite{https://doi.org/10.1002/ese3.661}). There also follows a mass and flow transferring process during the formation of contamination which is affected by the variation of hydrothermal parameters and physical properties of the contamination product (\cite{HE2018728}). This makes contamination development act as a physical process with a multivariable coupling influence. Furthermore, there lies a mechanism correlation between the upper and lower limits of the contamination length interval. How to clarify the internal association and incorporate it into the construction of mapping relationships by using mathematical expressions is another challenge (\cite{doi:10.1177/0144598720911158}).

In summary, the difficulty of this work can be concluded as follows:
\begin{enumerate}
    \item The complex multivariable coupling mechanisms and mechanism correlation characteristics of mixed oil formation cause a complex multi-dimensional feature space that is laborious to explore perfectly.
    \item The internal association between the upper and lower limit of contamination length interval which is influenced by the mass transfer process is an unclear mechanism correlation, and it is arduous to be represented by mathematical expressions.
\end{enumerate}

\subsection{Literature review}
\label{1.2}

Therefore, given these difficulties, it is necessary to conduct an investigation of the contamination formation and determine the evolution and development mechanism. To this begin, the early-stage research of mixed oil is conducted based on empirical models. \cite{smith1948interfacial} carried out a sequential transportation loop experiment to analyze the mixed oil formation and proposed the Smith \& Schulze formula. Then, \cite{sjenitzer1958much} considered the influence of friction on contamination formation and proposed Sjenitzer.F formula. In 1964, \cite{doi:10.1177/002034836317800160} analyzed the relation between the Reynolds number and contamination length and pointed out that the critical Reynolds number is only related to the inner diameter of the pipeline. Then, the Austin formula was proposed to calculate the contamination length, which also revealed the formation mechanism of contamination. By reviewing the aforementioned empirical models, it is indicated that the formation and development of contamination are related to the pipeline properties and hydraulic parameters (\cite{DU2022124689}). However, the empirical model neglects the effect of thermal factors and other parameters, leading to a weakness in model accuracy.

With the development of computer technology and computational fluid dynamics (CFD) technology, some scholars established multi-dimensional numerical models to analyze the formation process of contamination (\cite{LU2023105077}). In 2000, corrected the axial diffusion coefficient and proposed a novel two-dimensional mixed oil model. After that, \cite{10.1115/1.1459078} established a mixed oil calculation model based on Baptista's model by considering pipeline diameter, station operation, and hydraulic conditions. The numerical models provide a comprehensive analysis of contamination formation from the perspective of internal mechanisms (\cite{https://doi.org/10.1155/2019/6892915}). However, an unacceptable enormous cost occurs when the mixed oil numerical model for a long distance is required to be established. For example, solving the mixed oil migration in a 100 km pipeline takes several days (\cite{DU2023127452}).

The advancement of intelligent algorithms and computing power has witnessed a rise in machine learning (ML) models (\cite{DU2024129688}). ML models are applied to various industry problems (\cite{DU2023105647}), such as natural gas consumption prediction (\cite{DU2023125976}), pipeline shutdown pressure prediction (\cite{ZHENG2021518}), and power generation prediction (\cite{ZHENG2023113046}). \cite{DU2023127452} proposed a two-stage physics-informed neural network for predicting contamination concentration distribution. In the proposed model, the penalty terms of scientific theory are calculated by automatic differential and incorporated into the coupling loss function for accurate and explainable predictions. \cite{YUAN2023211466} proposed a physics-based Bayesian linear regression by defining the input features according to the Austin formula. Although these two methods made some progress in contamination length, the output variable is only the specific value of contamination length which cannot be used to determine the fluctuation interval. Furthermore, the multi-modal input features and different feature importance are neglected in these studies, and the gradient descent process lacks mechanism guidance, resulting to poor engineering applicability. \cite{DU2023128810} proposed a hybrid intelligent framework for prediction mixed oil concentration (MOC) but the contamination length fluctuation interval cannot be acquired.

Despite the progress achieved in previous studies, there still exist some limitations as shown below:
\begin{enumerate}
    \item The empirical models neglect the effect of hydrothermal factors and other parameters on mixed oil development, leading to a weakness in model accuracy.
    \item The conventional ML models lack comprehensive feature space construction and feature importance extraction based on different modal input features, and the gradient descent process lacks the guidance of contamination migration mechanism, thus reducing the model interpretability and engineering practicability.
    \item To the best of our knowledge, no study is designed to predict detailed fluctuation interval of contamination length from the perspective of practical expertise for developing more economical contamination cutting plans.
\end{enumerate}

\begin{figure}
    \centering
    \includegraphics[width=1\linewidth]{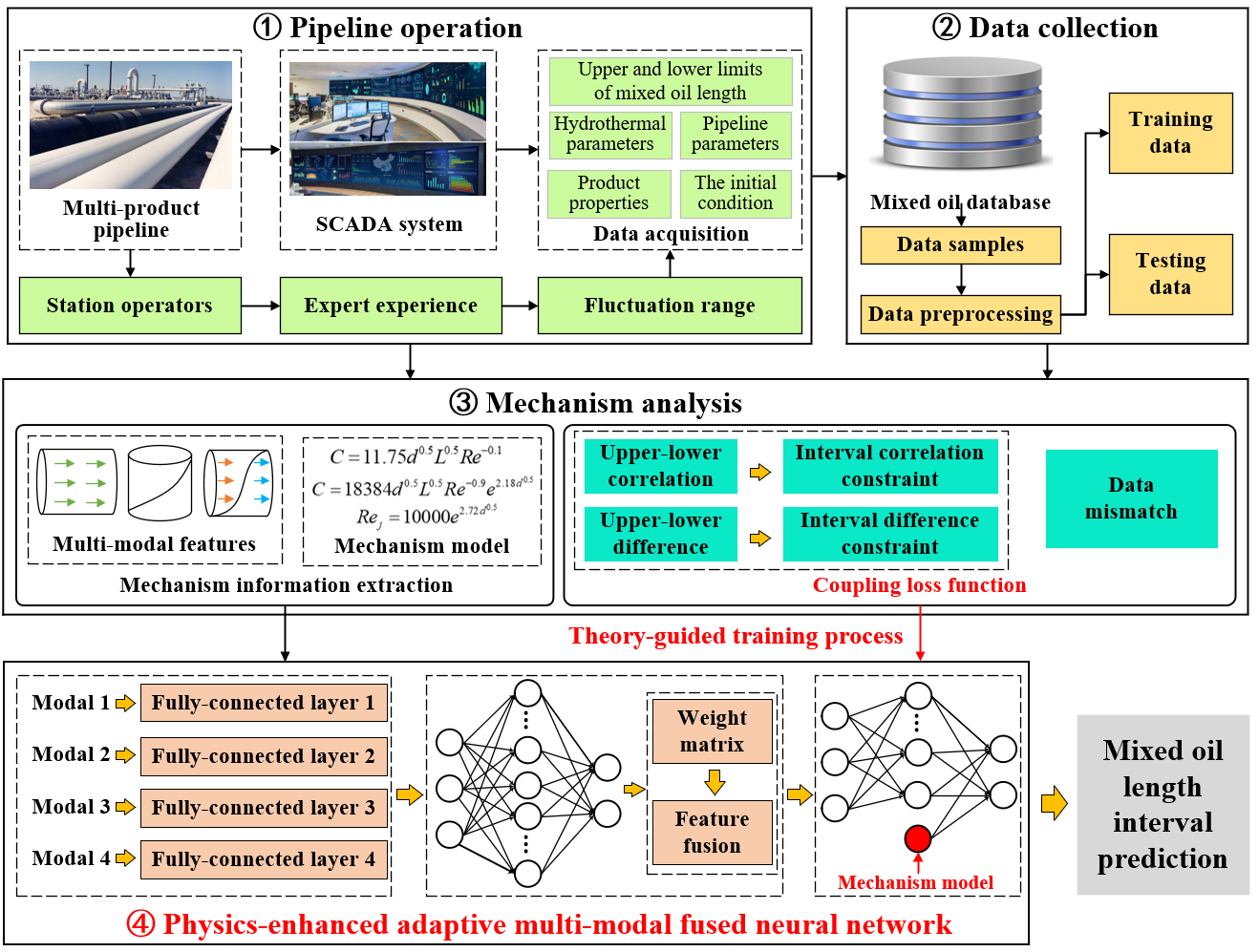}
    \caption{The research architecture of this study}
    \label{fig:1}
\end{figure}

To fill the gap of the previous studies, this work proposes a physics-enhanced adaptive multi-modal fused neural network (PE-AMFNN) to predict fluctuation interval of contamination length, as shown in Fig. \ref{fig:1}. To entrust a better practical application performance, the expert experience of station operators in pipeline control center is analyzed to determine the upper and lower limit of fluctuation interval. Noticing that a mass of operation data is stored in the Supervisory Control And Data Acquisition (SCADA) system of the multi-product pipeline, which is the center of pipeline operation monitoring. The operating data and contamination length data are collected from the SCADA system to form the contamination database. After analyzing the interaction relationship between various kinds of operating data and contamination length interval, an adaptive multi-modal feature fusion module is proposed to provide a holistic feature representation. Subsequently, it is noticed that the information forward propagation in the neural network lacks the guidance of mixing mechanism and the Austin formula possesses domain knowledge of mixed oil migration, so a mechanism coupled customized neural network is designed for model development. Furthermore, considering that there exists a mechanism correlation between the upper and lower limit of contamination length interval, a coupling loss function, which incorporates interval correlation and interval difference constraint is established to guide the backward propagation process and force the proposed model to obey the underlying theory. Subsequently, the proposed hybrid model, which is capable of learning complex multi-modal feature correlations and exploring the internal evolution mechanism of contamination is established to provide contamination length interval prediction with high-precise and physical consistency. This work can reduce the amount of contamination products, thus decreasing energy consumption.

\subsection{Contribution}
The contributions of this work can be summarized as follows:
\begin{enumerate}
    \item An adaptive multi-modal feature fusion strategy is proposed to capture the comprehensive correlation patterns between various kinds of input features and contamination length ranges, and further quantify different influencing mechanisms.
    \item A mechanism-coupled customized neural network is designed to incorporate migration mechanism into the forward and backward propagation process, then, a coupling loss function is established to force the neural network to conform to the explicit knowledge of contamination length interval, thus improving the interpretability and effectiveness of the proposed model.
    \item As far as we know, the first physics-enhanced multi-modal fused modeling paradigm for contamination length range prediction from the perspective of practical expertise is proposed for more interpretable and practical useful contamination development information.
    \item Real-world cases of contamination length range in multi-product pipelines are utilized to testify to the effectiveness and accuracy of the proposed model, and the sensitivity analysis is conducted to illustrate the significance of different modules.
\end{enumerate}

The rest of this paper is organized as follows. Section \ref{section2} explains the problem formulation of contamination length range prediction and elaborates on the principle of the proposed model in detail. Section \ref{section3} uses several real-world cases of contamination length range to verify the accuracy of the proposed model and highlight the significance of each module by a sensitivity analysis. Eventually, the conclusion and future work are discussed in Section \ref{section4}.

\section{Physics-enhanced adaptive multi-modal fused neural network}
\label{section2}

\begin{figure}
    \centering
    \includegraphics[width=1\linewidth]{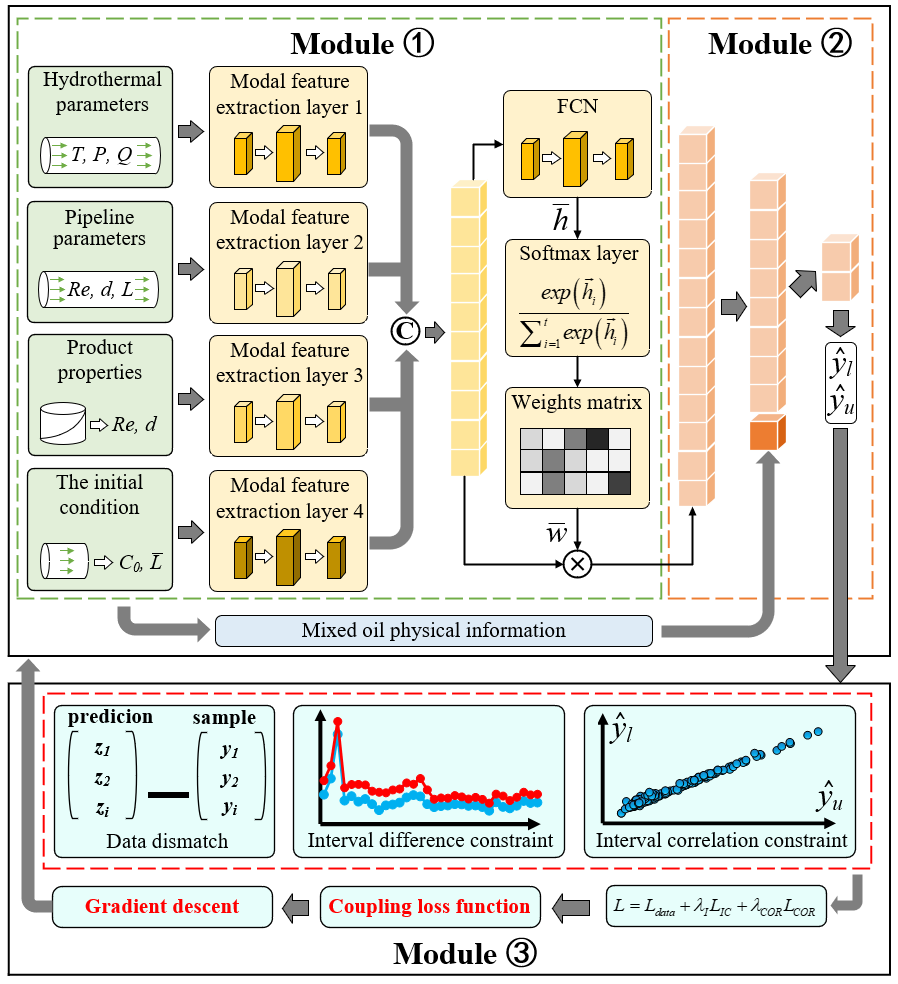}
    \caption{The framework of PE-AMFNN}
    \label{fig:2}
\end{figure}

As depicted in Fig. \ref{fig:2}, a novel hybrid intelligent framework, PE-AMFNN, is designed to predict contamination length interval. The proposed PE-AMFNN is composed of three functional modules, as shown below:
\begin{enumerate}
    \item An adaptive multi-modal feature fusion module (Module 1), which consists of several multi-modal feature extraction layers and adaptive weight learning, is developed to capture the correlation patterns of multi-modal variables and weights each variable according to their feature importance.
    \item A mechanism-coupled customized neural network (Module 2), which applies the mechanism model result as a neural node to participate in the forward and backward propagation process, is proposed to make the data-driven function and physical knowledge cooperate and improve the interpretability and accuracy of PE-AMFNN.
    \item A physics-embedded coupling loss function (Module 3) integrates interval difference and interval correlation constraint into the training process to guide the backward propagation process, aiming to enforce the neural network drawing physically reasonable results thus improving the effectiveness of PE-AMFNN.
\end{enumerate}

\subsection{Module 1: Adaptive multi-modal feature fusion}

As a complex flow and mass transfer process, the mixed oil migration is affected by multiple factors. In our previous work, the development mechanism of mixed oil migration is analyzed thoroughly to determine the influencing variables. Although a comprehensive feature subset of mixed oil migration is figured out and established, the correlation patterns and latent feature representations of multiple variables are neglected and unclear in neural networks. Furthermore, the multi-modal information possesses various degrees of impact on mixed oil migration and different feature importance, which also lacks sufficient reflection in previous work.
The reference \cite{DU2023128810} analyzed feature variables and concluded the factors that influence contamination length. It is worth noting that, the detailed description of the variables in Table 1 as well as other variables in the following content can be found in Nomenclature. According to the previous studies, it is indicated that there lies different correlation mechanisms between input features and contamination length. For example, the hydrothermal parameters mainly determine the contamination length by affecting the diffusion coefficient and the pipeline parameters have an impact on the contamination length by determining the thickness of the boundary layer. Additionally, the diffusion intention and initial degree of mixed oil development are affected by product properties and the initial condition, thus causing the change of contamination length. Consequently, it is of great necessity and significance to extract latent representations of different modal features.

\begin{figure}
    \centering
    \includegraphics[width=1\linewidth]{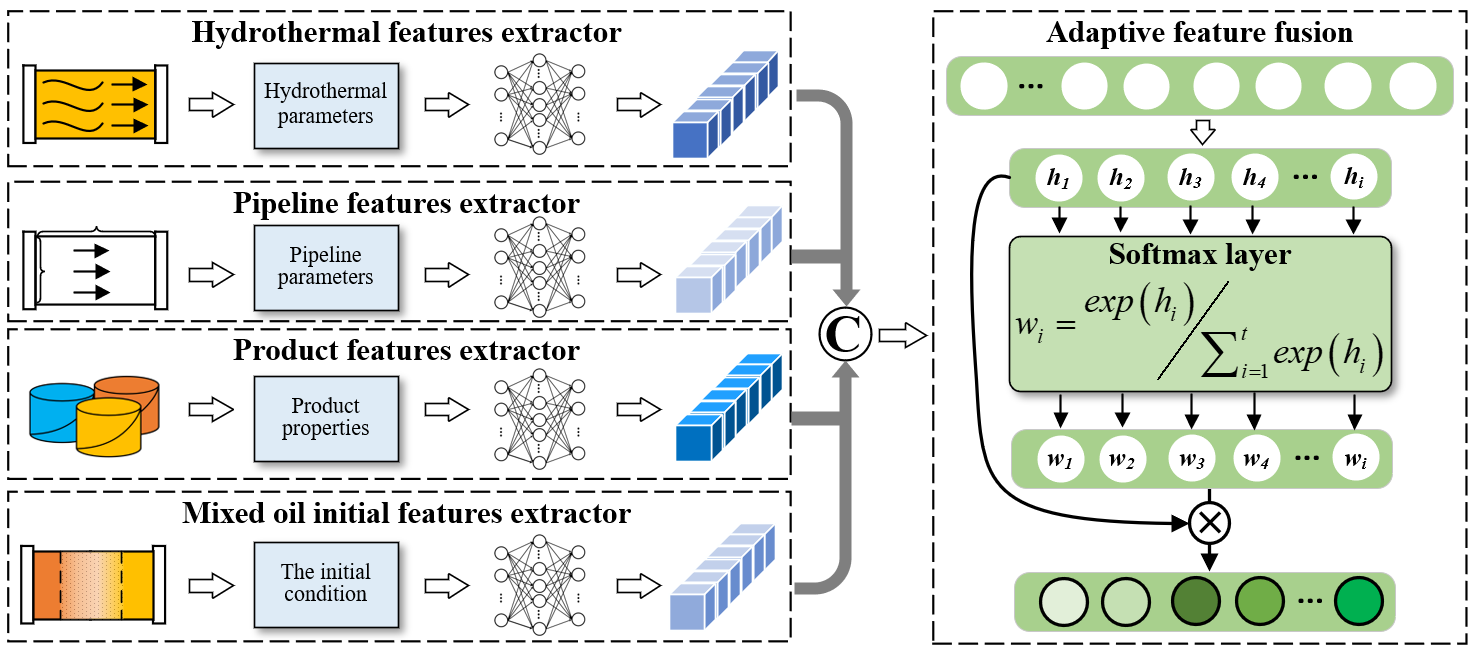}
    \caption{The architecture of Module 1}
    \label{fig:3}
\end{figure}

As shown in Fig. \ref{fig:3}, several multi-modal feature representation extractors are established which are capable of capturing private features from each modal data. Among them, the input features are first fed into the fully connected network (FCN), which is designed to extract latent representations and correlation patterns of multi-modal features, aiming to obtain the private feature vectors and interaction information. To provide a better ability for feature extraction, the architecture of FCN for each modal data is optimized by trial and error and may not be the same as others. After that, the feature vectors of each modal data are concatenated together to form a holistic feature representation.

Subsequently, an adaptive feature fusion framework is designed to assign different weights to feature vectors, intending to consider the feature importance. The concatenated feature representation is passed through a SoftMax function layer to acquire the corresponding weights of each feature. To reflect the importance of the feature extraction process, the feature weights are element-wise multiplied with feature nodes to obtain the new feature representation with feature importance. Among them, the weighting process of feature nodes is self-adaptive during the training process, aiming to provide a more comprehensive feature representation. The mathematical formula of the adaptive feature fusion process is depicted below:

\begin{equation}\label{eqn-1}
w_i=\frac{exp(h_i)}{\sum_{i=1}^texp(h_i)}
\end{equation}

\begin{equation}\label{eqn-2}
V_{_w}=V\otimes W
\end{equation}

where $h_{i}$ represents the feature nodes in concatenated feature representation and the $w_{i}$ is the feature weights. $W$ denotes the weight matrix consisting of feature weights and $V_{w}$ is the weighted feature representation. $V$ is the fused holistic feature representation.

\subsection{Module 2: Mechanism-coupled customized neural network}
\label{2.2}

\begin{figure}
    \centering
    \includegraphics[width=1\linewidth]{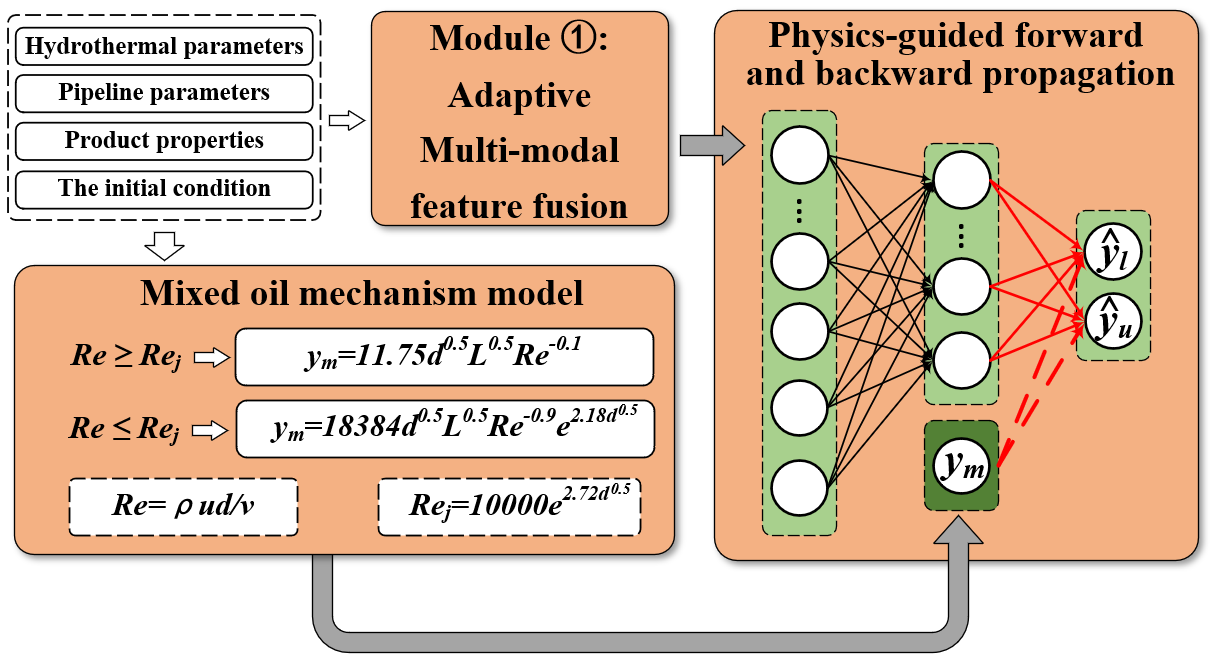}
    \caption{The architecture of the mechanism-coupled customized neural network}
    \label{fig:4}
\end{figure}

After acquiring the weighted feature representation, the feature information will be passed to the following FCN to approximate the nonlinear function between contamination length interval and feature variables. Nevertheless, it is suggested that the conventional FCN possesses poor ability for capturing potential knowledge of the mechanism evolution process even though retaining excellent nonlinear approximating function. This issue causes a tough problem in that the proposed model is provided with adequate capability of interpretability and accuracy, further limiting the application potential in pipeline engineering practice.
To this beginning, this work proposes a mechanism-coupled customized neural network by integrating physical information in a mixed oil mechanism model into the forward and backward propagation process of the neural network, as depicted in Fig. \ref{fig:4}. From the discussion in Section \ref{1.2}, it is indicated that the empirical model, such as Austin formula, reveals inherent principles of mixed oil migration to a certain extent though possesses large deviations. Inspired by the information transmission process of the neural network, a customized neural network that adds mechanism model results as a neural node is designed.
Austin formula result announces the correlation between mixed oil length and several feature variables. The calculation process of Austin formula is shown below:

\begin{equation}\label{eqn-3}
\begin{cases}y_m=11.75d^{0.5}L^{0.5}Re^{-0.1}&Re\geq Re_j\\y_m=18384d^{0.5}L^{0.5}Re^{-0.9}e^{2.18d^{0.5}}&Re\leq Re_j\end{cases}
\end{equation}
\begin{equation}\label{eqn-4}
Re_j=10000e^{2.72d^{0.5}}
\end{equation}
where $d$ and $L$ are the diameter and length of the multi-product pipeline (m). $Re_{j}$ and $Re$ are the critical Reynolds number and Reynolds number, respectively. $y_{m}$ is the mixed oil length (m).

The output of Austin formula is used as a new neural node added into the hidden layer, as depicted in Fig. \ref{fig:4}. The prediction results are informed about not only the relationship between input and output variables but also the physical potential information of mixed oil migration. The physical potential information contributes to the learning process of neural networks. During the training process of weights and bias, the neural network can learn how to manipulate the nonlinear function according to the data-driven based latent representation and mixed oil migration mechanism, thus the better match results between the predictions of the proposed model and the observed results. In other words, the output of Austin formula can participate in the forward and backward propagation process and guide the update of weights and bias in the neural network, intending to improve the accuracy and interpretability. The mathematical formula of the information transfer process in the mechanism-coupled customized neural network is shown below:

\begin{equation}\label{eqn-5}
\begin{pmatrix}\hat{y}_l,\hat{y}_u\end{pmatrix}=f\begin{pmatrix}w_dx_d+w_py_m+b\end{pmatrix}
\end{equation}

where $w_{d}$ and $w_{p}$ are the weight matrix of data-driven network nodes and physics-guided network node.  $b$ is the bias in the customized network layer.  $f(\bullet)$denotes the nonlinear activation function.

\subsection{Module 3: Physics-embedded coupling loss function}

The contamination length interval prediction is essentially a multiple-output prediction problem with strong mechanism correlation patterns. This brings several challenges for model development. Firstly, gathering and annotating operating data in a multi-product pipeline to form the contamination length interval dataset is time-consuming and labor-intensive. So, using a limited amount of data to develop the prediction model is a common issue, which easily leads to the degeneration of model effectiveness and accuracy. Secondly, there are strong interaction relationship between these two predicted lower and upper limits. For example, the upper limit is larger than the lower limit by a migration length which can be obtained according to the distribution flowrate and the fixed operation time. However, the distribution flowrate is unknown before the mixed oil segment arrives at stations. Therefore, the precise deviation between these two limits is unclear but the minimum distribution flow rate in the operation plan can be used to calculate an approximative deviation. In addition, the lower and upper limits of contamination length interval are essentially the mixed oil migration results in the same pipeline which are influenced by similar variables. This fact implies that there must exist a statistical relationship between the lower and upper limits.

In module 2, a physic-guided forward and backward propagation process is designed, yet it cannot incorporate the correlation patterns between output results into model improvement. To overcome the aforementioned challenges, a physics-embedded coupling loss function is constructed to constrain the training process of the neural network and further avoid the degeneracy of the model. Considering that, the upper limit ought to be larger than the lower limit and an approximate gap between them can be determined by the operation plan, a difference constraint can be built, as shown below:

\begin{equation}\label{eqn-6}
DC=NN_l(X;\theta)+y_{IG}-NN_u(X;\theta)
\end{equation}

\begin{equation}\label{eqn-7}
y_{IG}=\frac{Q_m\bullet T_d}{600}
\end{equation}

where $y_{IG}$ is the length range gap (m). $Q_{m}$ denotes the minimum distribution flowrate (m\textsuperscript{3}/h ). $T_{d}$ are the delay duration (min). $NN(X;\theta)$ is the predicted results from the neural network. It is worth noting that the contamination range gap calculated by the minimum distribution flowrate is always smaller than the true gap, so the value of the difference constraint is respected to be negative. To reflect the contamination range difference constraint in the training process, a penalty function is constructed as shown below:

\begin{equation}\label{eqn-8}
f_{DC}=\text{Re}\operatorname{LU}\Big[NN_l(X;\theta)+y_{IG}-NN_u(X;\theta)\Big]
\end{equation}

\begin{equation}\label{eqn-9}
L_{DC}=\frac{1}{N}\sum_{i}\left|\mathrm{Re}\operatorname{LU}\Big[NN_{l}(X_{i};\theta)+y_{IG}^{i}-NN_{u}(X_{i};\theta)\Big]\right|^{2}
\end{equation}

where $ReLU(\bullet)$ is the rectified linear unit function, which outputs zero if the input is negative and returns the original value when the input is positive. $L_{DC}$ represents the loss term corresponding to the interval difference constraint.

Furthermore, to use the statistical correlation to guide the training process, a linear correlation formula (Eq \ref{eqn-10}) is resumed and established in this work. In the range correlation constraint module, the correlation weights ($w_{c}$, $b_{c}$) are defined and updated during the training process to obtain the linear transformation results of lower limits. To constrain the original distribution of the upper limit and the correlated feature distribution of the lower limit to be as close as possible, an range correlation constraint is proposed and the penalty function (Eq \ref{eqn-12}) is constructed to force the difference between these two distributions smaller.

\begin{equation}\label{eqn-10}
g_l=w_c\bullet NN_l\left(X;\theta\right)+b_c
\end{equation}

\begin{equation}\label{eqn-11}
f_{COR}=P\Big[w_c\bullet NN_l\left(X;\theta\right)+b_c\Big]log\frac{P\Big[w_c\bullet NN_l\left(X;\theta\right)+b_c\Big]}{Q\Big[NN_u\left(X;\theta\right)\Big]}
\end{equation}

\begin{equation}\label{eqn-12}
L_{COR}=\frac{1}{N}\sum_{i}\left|P\Big[w_{c}\bullet NN_{l}\left(X_{i};\theta\right)+b_{c}\Big]log\frac{P\Big[w_{c}\bullet NN_{l}\left(X_{i};\theta\right)+b_{c}\Big]\Big]}{Q\Big[NN_{u}\left(X_{i};\theta\right)\Big]}\right|
\end{equation}

where $P(\bullet)$ and $Q(\bullet)$ are the probability distribution of input data. The lower $L_{COR}$ suggests a stronger linear correlation between the lower and upper limit. In this way, a physic-embedded loss function (Eq \ref{eqn-13}) can be established to guide the backward propagation process of the neural network, which is expected to provide a more interpretable and accurate prediction model. Among this, the $\lambda_{DC}$ and $\lambda_{COR}$ are the hyper-parameters of the coupling loss function, which is used to determine the weights of each penalty and can be adjusted during the training process to ensure a better performance of the proposed model. $L_{data}$ is the mean square error between predicted results and observed results.

\begin{equation}\label{eqn-13}
L\begin{pmatrix}\theta\end{pmatrix}=L_{data}+\lambda_{DC}L_{DC}+\lambda_{COR}L_{COR}
\end{equation}
After the coupling loss function is converged, the proposed model is trained and the predicted contamination length interval can be used to assist the contamination reception.

\section{Results and discussion}
\label{section3}

In this section, several cases of real-world multi-product pipelines are taken as examples to validate the effectiveness and accuracy of the proposed model. To assess the overall performance of the proposed model, several advanced models are utilized as comparative models. Eventually, a sensitivity analysis is carried out to illustrate the importance of model components.

\subsection{Data description}

To provide reliable comparison results of various prediction models, an available dataset with practical contamination length samples in a multi-product pipeline ought to be established. In this study, the pipeline operation data in the SCADA system of a multi-product pipeline network in China is gathered. Specifically, the operation data from March 14, 2020, to April 14, 2020, and October 14, 2020, to November 14, 2020, are collected. Then, the original data samples are preprocessed to remove the outliers and missing values. Some of the features are transformed to acquire the input feature as mentioned in Section \ref{section2}. In this way, the contamination length database, which contains 350 samples, can be used for further experiments.

To ensure the robustness and effectiveness of the experimental results, the data division ought to be conducted to carry out the cross-validation. In this study, the original dataset is randomly shuffled and split into three different datasets by the K-fold method, namely, 80 \% of them are taken as the training set for adjusting the weights and bias in the prediction models, 10 \% of them are taken as the validation set for verifying the performance after each training epoch is finished, and 10 \% of them are taken as the testing set for testifying to the accuracy of the prediction model.

\subsection{Experimental setting}

The parameter setting of the data-driven model possesses a significant impact on model performance. The hyper-parameters of the proposed model consist of three different parts, including Module 1, Module 2, and Module 3. By trial and error, the suitable hyper-parameters in these three sections are determined. Each multi-modal extraction layer only contains one hidden layer. The number of neural nodes in the hydrothermal parameters extractor, pipeline parameters extractor, product properties extractor, the initial condition extractor are set as 12, 8, 10, 5, respectively. The number of neural nodes in the FCN layer is set as 20 to fuse different multi-modal information. Besides, the dropout is set as 0.1 to avoid overfitting.

For Module 2, the number of layers is set as 2, and the number of neural nodes is selected as [60, 10]. In Module 3, the hyper-parameters that influence the prediction accuracy and effectiveness are the weights of penalty terms. By trial and error, the values of $\lambda_{DC}$ and $\lambda_{COR}$ are set as 0.01, and 0.1 to obtain a better convergence performance.

Last but not least, the activation function of the proposed model is set as ReLU to improve the nonlinear approximation function. The proposed model is trained with the MSE loss function, and the Adam optimizer is adopted. The proposed model is trained for 2000 rounds, and the initial learning rate is set as 0.0001. Furthermore, all the deep learning models are executed on the Pytorch framework for model establishment.

\subsection{Validation and comparison of the proposed model}

After determining the suitable structure of the proposed model and the available data, comprehensive comparisons are conducted to demonstrate the accuracy of the proposed model. The experiment of each prediction model is conducted ten times to acquire the average values of evaluation metrics. To this begin, several start-of-art models, such as the Decision Tree Regressor (DTR), K-Nearest Neighbor Regressor (KNR), Extra Tree Regressor (ETR), Random Forest Regressor (RFR), and Deep Neural Network (DNN) serve as the comparative models. Additionally, several recent achievements (\cite{YUAN2023211466} and \cite{CHEN2021108787}) in contamination length prediction are applied for advancement assessment. It is worth noting that the suitable hyper-parameters of the comparative models are adjusted to obtain the optimal prediction results.

\begin{figure}[ht]
    \centering
    \begin{subfigure}[b]{0.4\textwidth}
        \includegraphics[width=\textwidth]{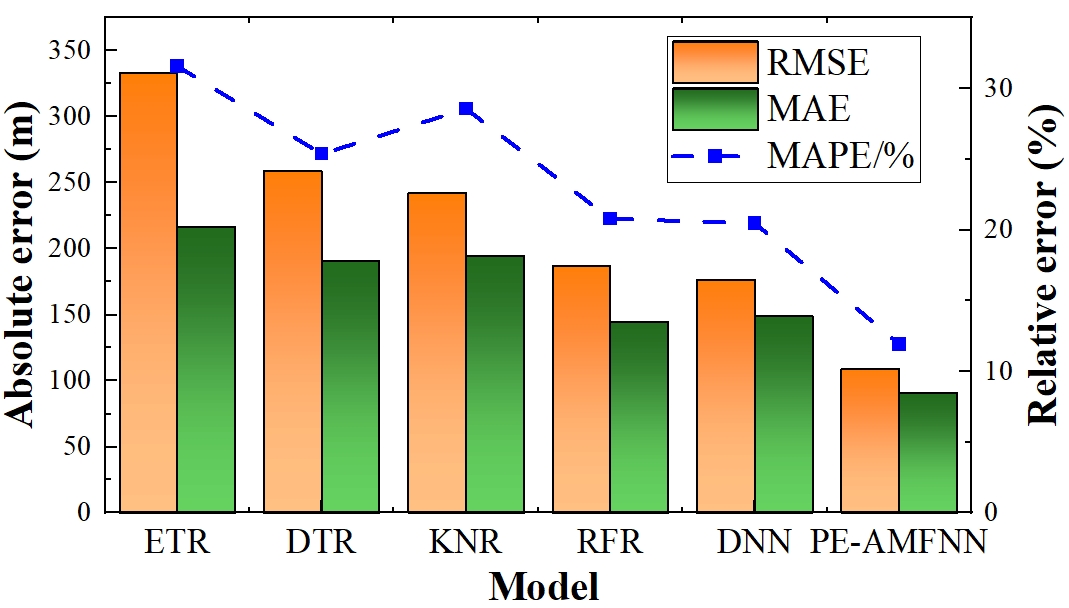}
        \caption{Lower limit prediction}
        \label{fig:sub1}
    \end{subfigure}
    \hfill % 用于两张子图之间添加空格
    \begin{subfigure}[b]{0.4\textwidth}
        \includegraphics[width=\textwidth]{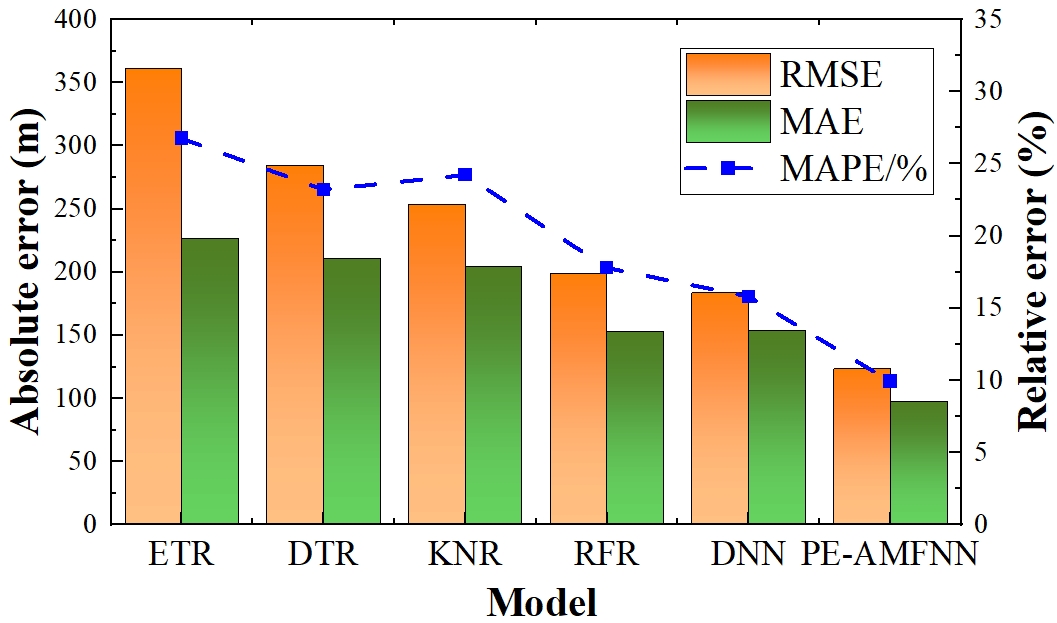}
        \caption{Upper limit prediction}
        \label{fig:sub2}
    \end{subfigure}
    \caption{The comparison of state-of-art models and the proposed model}
    \label{fig:5}
\end{figure}

Fig. \ref{fig:5} depicts the prediction error comparisons between the proposed model and the start-of-art models. Apparently, the proposed model acquires more accurate prediction results than that of other comparative models, with the lowest prediction errors. From the comparison of the proposed model and DNN, it is indicated that carrying out the multi-modal adaptive feature fusion and using the physical information of contamination development to guide the forward and backward propagation process is of importance for accuracy improvement. The proposed model achieves a decrease of RMSE, MAE, and MAPE by 43\%, 39\%, and 45\% than that of the DNN model in lower limit prediction. In contrast, the conventional regression tree-based machine learning models obtain worse prediction performance. For example, the ETR model yields the prediction results with less accuracy, and the RMSE is increased by 204\% compared to PE-AMFNN in upper limit prediction. This fact implies that the ETR model is not suitable for contamination length interval prediction. Conversely, the ensemble machine learning model such as the RFR achieves better performance with RMSE and MAPE being 187 m and 20\% in lower limit prediction. It is indicated that the ensemble learning approach allows more efficient nonlinear correlation learning. Overall, the proposed hybrid prediction model realizes the best performance in contamination length interval prediction.
 
\begin{figure}[ht]
    \centering
    \begin{subfigure}[b]{0.4\textwidth}
        \includegraphics[width=\textwidth]{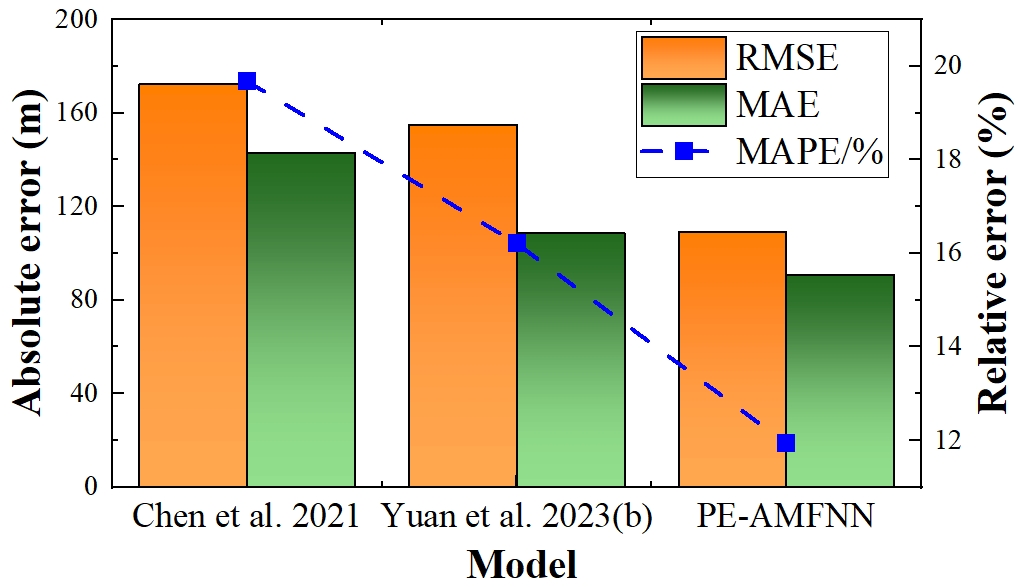}
        \caption{Lower limit prediction}
        \label{fig:sub3}
    \end{subfigure}
    \hfill % 用于两张子图之间添加空格
    \begin{subfigure}[b]{0.4\textwidth}
        \includegraphics[width=\textwidth]{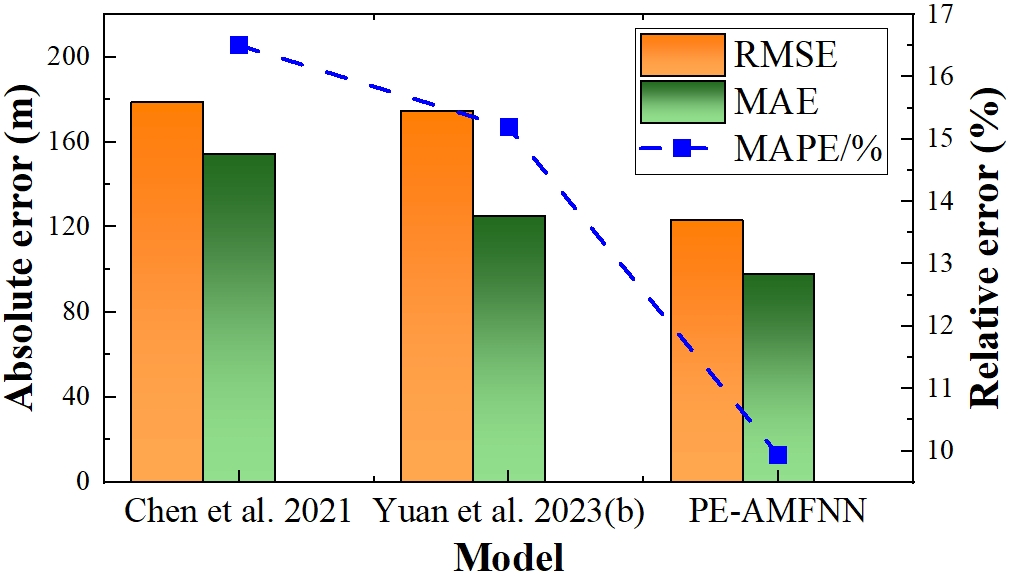}
        \caption{Upper limit prediction}
        \label{fig:sub4}
    \end{subfigure}
    \caption{The comparisons of the recent achievements and the proposed model}
    \label{fig:6}
\end{figure}

Only the comparisons between the proposed model and the start-of-art models cannot provide an evident conclusion. For this reason, the comparisons of prediction errors between the latest achievements and the proposed model are conducted, as shown in Fig. \ref{fig:6}. It is suggested that PE-AMFNN achieves the best performance in contamination length interval prediction with the lowest absolute and relative errors, followed by \cite{YUAN2023211466} and \cite{CHEN2021108787}. The model proposed in \cite{CHEN2021108787} improves the input features by analyzing the mixed oil mechanism formula and predicts the contamination length by using an ensemble tree algorithm. However, the prediction errors are higher than that of the model proposed in \cite{YUAN2023211466}, potentially attributed to the inability of the prediction model to approximate the nonlinear correlations compared to the physics-based Bayesian model. Even though, the model in \cite{YUAN2023211466} only acquires a slight reduction compared to \cite{CHEN2021108787}, with RMSE, MAE, and MAPE reducing 10\%, 24\%, and 16\% in lower limit prediction. This may result from the lack of comprehensive feature correlations and physical information guidance. In contrast, the proposed model achieves a prominent improvement of model performance compared to the latest achievements, with RMSE, MAE and MAPE reduced by 30\%, 17\%, and 31\%. Same conclusions can be drawn in upper limit prediction (Fig. \ref{fig:sub4}). This fact proves that the extraction of more holistic feature correlations and the integration of physical principle in forward and backward propagation process are extremely important for model improvement.

\subsection{Mixed oil length interval prediction}

\begin{figure}[ht]
  \centering
  % 上面的子图
  \begin{subfigure}[b]{0.8\textwidth}
    \includegraphics[width=\textwidth]{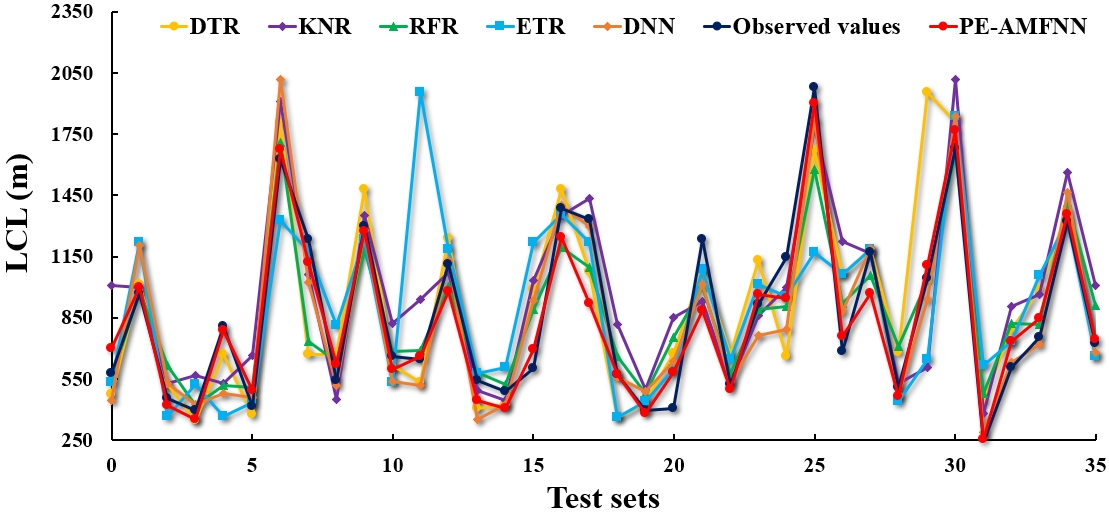}
    \caption{Lower limit prediction}
    \label{fig:upper1}
  \end{subfigure}
  
  % 下面的子图
  \begin{subfigure}[b]{0.8\textwidth}
    \includegraphics[width=\textwidth]{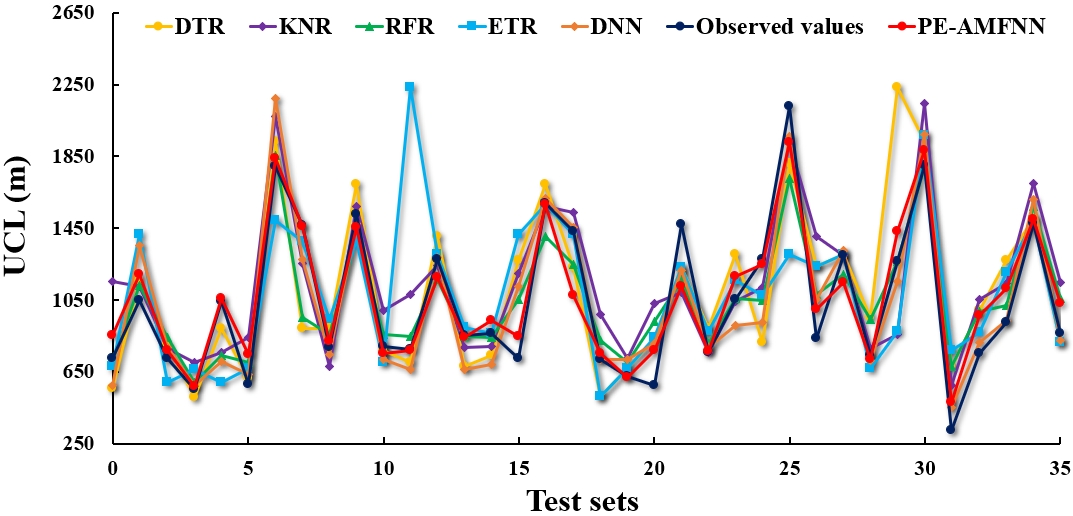}
    \caption{Upper limit prediction}
    \label{fig:lower1}
  \end{subfigure}
  \caption{The prediction results of state-of-art models and the proposed model}
  \label{fig:7}
\end{figure}

\begin{figure}[ht]
  \centering
  % 上面的子图
  \begin{subfigure}[b]{0.8\textwidth}
    \includegraphics[width=\textwidth]{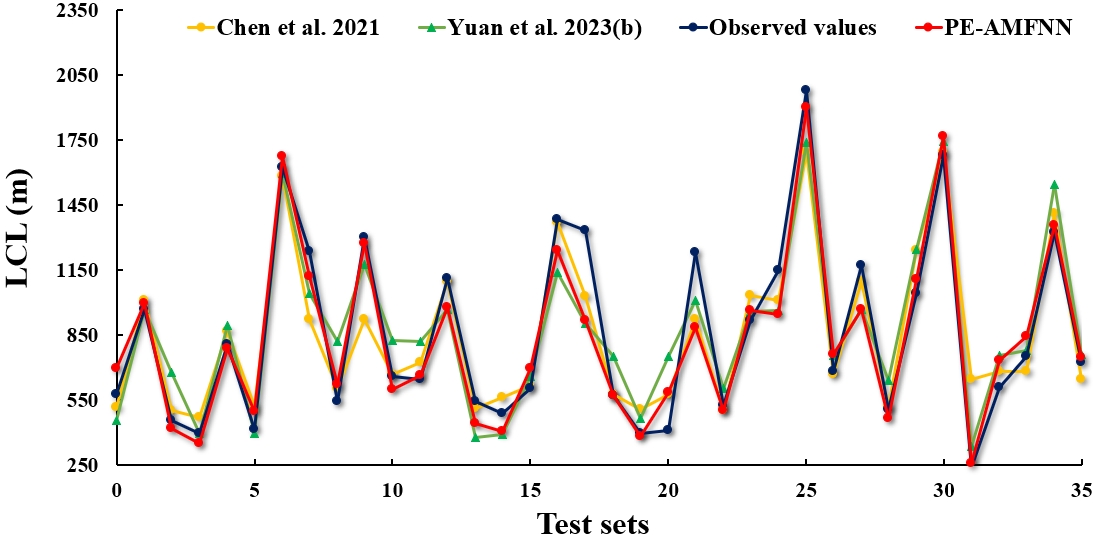}
    \caption{Lower limit prediction}
    \label{fig:upper2}
  \end{subfigure}
  
  % 下面的子图
  \begin{subfigure}[b]{0.8\textwidth}
    \includegraphics[width=\textwidth]{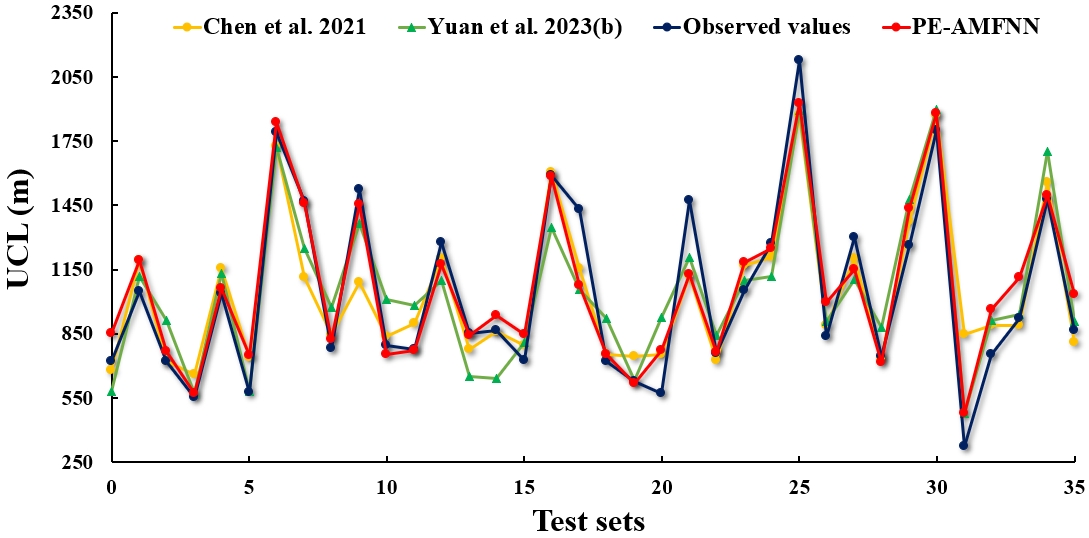}
    \caption{Upper limit prediction}
    \label{fig:lower2}
  \end{subfigure}
  \caption{The prediction results of exciting methods and the proposed model}
  \label{fig:8}
\end{figure}

Fig. \ref{fig:7} and \ref{fig:8} depict the comparisons of predicted curves and observed curves for various models. The black line represents the observed values of the lower and upper limits. Apparently, the results predicted by the proposed model are closest to the observed values and achieve the lowest deviations compared to other models, reflecting its better prediction capability. Significant deviations appear in the results predicted by the conventional machine learning algorithms, including DTR, and KNR. Benefiting the ensemble learning enhancement, the RFR model reduces the predicted errors compared to DTR and KNR. The DNN model applies several hidden layers to capture the nonlinear interaction relationship in multiple feature variables, thereby obtaining more exact prediction performance. From the comparisons between previous studies and the proposed model in Fig. \ref{fig:8}, the prominent strength in contamination length interval prediction of the proposed model is embodied. From the fifth test point to the fifteenth test point in Fig. \ref{fig:8}, the previous studies acquire higher deviations from the observed curves and the predicted curves are far away from the observed curves. By adaptively extracting multi-modal feature correlations and physical principles, the proposed model draws predicted results with the slightest deviation and the predicted curves fit the observed curves better.

\subsection{Sensitivity analysis of model components}

\begin{figure}[ht]
  \centering
  % 上面的子图
  \begin{subfigure}[b]{0.8\textwidth}
    \includegraphics[width=\textwidth]{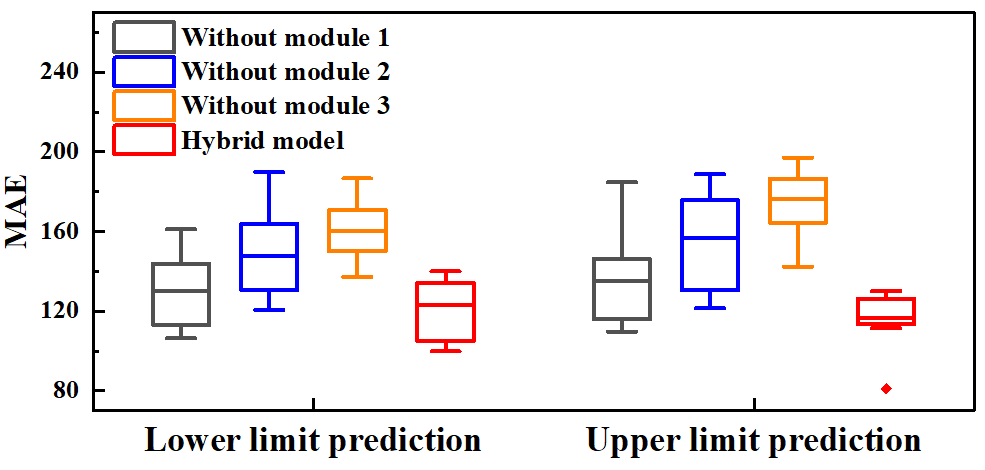}
    \caption{MAE}
    \label{fig:upper3}
  \end{subfigure}
  
  % 下面的子图
  \begin{subfigure}[b]{0.8\textwidth}
    \includegraphics[width=\textwidth]{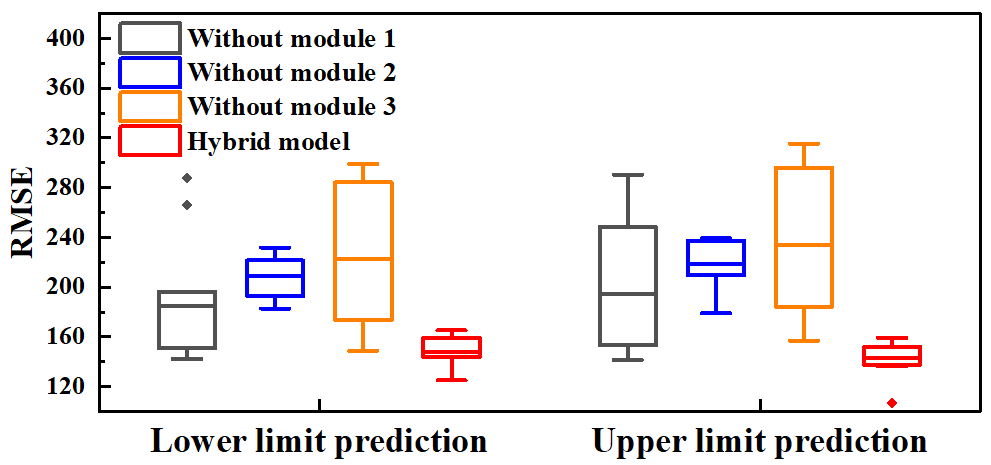}
    \caption{RMSE}
    \label{fig:lower3}
  \end{subfigure}
  \caption{The sensitivity analysis of different model components}
  \label{fig:9}
\end{figure}

The previous content elaborates on the superiority of the proposed model compared to the start-of-art models and latest studies. To make an in-depth analysis of the hybrid model, this subsection carries out a sensitivity analysis of different functional modules to illustrate the significance of model development. As depicted in Fig. \ref{fig:9}, the box plots are utilized to compare the prediction errors. As introduced in Section \ref{section2}, the hybrid intelligent model is made up of three different modules, including a multi-modal adaptive feature extraction module, a mechanism-coupled customized neural network module, and a physics-embedded coupling loss function module. To set up a control experiment, each different model module in the hybrid model is removed to construct different comparative models, which are the model without module 1, the model without module 2, and the model without module 3.
As shown in Fig. \ref{fig:9}, the predictions of different comparative models are repeated ten times to obtain the median, lower quartile, upper quartile, maximum values, and minimum values. Obviously, the model without module 3, which is the model only using the MSE loss of data mismatch to guide the training process, acquires the highest prediction errors and worst prediction performance. This phenomenon implies a prominent performance improvement on the prediction model by integrating physics-embedded loss function into the training process to guide the backward propagation. By comparing the results predicted by the model without module 2 and the model without module 3, it is indicated that the incorporation of a mechanism-coupled customized neural network shows less significance than the physics-embedded coupling loss function. By removing the multi-modal adaptive feature extraction module from the hybrid model, the prediction errors are higher than that of the proposed model but lower than that of the model without module 2. It is proved that the incorporation of the mechanism principle is more important for performance improvement than multi-modal feature fusion. However, the prediction results represent significant deviations if any module is removed from the proposed model. To this end, it is concluded that the physical-embedded loss function for backward propagation represents a core effect among all three functional modules in model development. Furthermore, incorporating multi-modal adaptive feature fusion and physical principles into the model establishment is crucial for a more accurate and effective prediction result.

\section{Conclusion}
\label{section4}

In this work, a hybrid intelligent model that integrates multi-modal feature fusion and physical principles is proposed to address the insufficient feature representation and physically unreasonable prediction results of conventional prediction methods. The main innovations and advantages of the proposed model can be summarized as follows::
\begin{enumerate}
    \item To overcome the shortcomings of imperfect feature representation, a multi-modal adaptive features fusion module, which contains a multi-variables extraction component and an adaptive feature fusion component, is designed for a more comprehensive latent feature correlations acquirement. The proposed module can establish a more holistic feature space and quantify the impact degree of different feature variables, hence improving the accuracy and generalization.
    \item To avoid the deficiency of mechanism information extraction capability, a mechanism-coupled customized network layer is designed for collaborative guidance on mechanism information and nonlinear correlations, which incorporates the output results of the mechanism model into the forward and backward propagation process. This module can apply the mechanism formula to guide the model prediction, thus enhancing the accuracy and interpretability.
    \item To address the absence of physical principles in the training process, a physics-embedded loss function, which creates an interval difference penalty and an interval correlation penalty according to the domain knowledge of contamination length, is established to enforce the model drawing physically unreasonable results. This module can restrict the backward propagation of neural networks, thereby increasing the interpretability and robustness.

\end{enumerate}
Evaluation of real-world contamination length cases proves that the proposed model is capable of providing more accurate and effective prediction results, with RMSE, MAE, and MAPE reducing by 67\%, 58\%, and 65\%, compared to conventional machine learning models. The sensitivity analysis also proves an important model improvement by feature fusion and physical principles integration. The limitation of this work is that it is difficult to acquire the most suitable weights of different penalty terms in the coupling loss function, which has a significant impact on model performance. Therefore, our future work will mainly focus on the method of determining the weights of penalty terms for the optimal model performance.

{
\small

\bibliographystyle{plainnat}
\bibliography{cites/1,cites/2,cites/3,cites/4,cites/5,cites/6,cites/7,cites/8,cites/9,cites/10,cites/11,cites/12,cites/13,cites/14,cites/15,cites/16,cites/17,cites/18,cites/19,cites/21,cites/22,cites/23,cites/24,cites/25,cites/26,cites/27,cites/28,cites/29,cites/47}

\begin{thebibliography}{29}
\providecommand{\natexlab}[1]{#1}
\providecommand{\url}[1]{\texttt{#1}}
\expandafter\ifx\csname urlstyle\endcsname\relax
  \providecommand{\doi}[1]{doi: #1}\else
  \providecommand{\doi}{doi: \begingroup \urlstyle{rm}\Url}\fi

\bibitem[Abdellaoui et~al.(2021)Abdellaoui, Souier, Sahnoun, and {Ben Abdelaziz}]{ABDELLAOUI2021107483}
Wassila Abdellaoui, Mehdi Souier, M’hammed Sahnoun, and Fouad {Ben Abdelaziz}.
\newblock Multi-period optimal schedule of a multi-product pipeline: A case study in algeria.
\newblock \emph{Computers \& Industrial Engineering}, 159:\penalty0 107483, 2021.
\newblock ISSN 0360-8352.
\newblock \doi{https://doi.org/10.1016/j.cie.2021.107483}.
\newblock URL \url{https://www.sciencedirect.com/science/article/pii/S0360835221003879}.

\bibitem[Austin and Palfrey(1963)]{doi:10.1177/002034836317800160}
J.~E. Austin and J.~R. Palfrey.
\newblock Mixing of miscible but dissimilar liquids in serial flow in a pipeline.
\newblock \emph{Proceedings of the Institution of Mechanical Engineers}, 178\penalty0 (1):\penalty0 377--389, 1963.
\newblock \doi{10.1177/002034836317800160}.
\newblock URL \url{https://doi.org/10.1177/002034836317800160}.

\bibitem[Bamoumen et~al.(2023)Bamoumen, Elfirdoussi, Ren, and Tchernev]{BAMOUMEN2023106082}
Meryem Bamoumen, Selwa Elfirdoussi, Libo Ren, and Nikolay Tchernev.
\newblock An efficient grasp-like algorithm for the multi-product straight pipeline scheduling problem.
\newblock \emph{Computers \& Operations Research}, 150:\penalty0 106082, 2023.
\newblock ISSN 0305-0548.
\newblock \doi{https://doi.org/10.1016/j.cor.2022.106082}.
\newblock URL \url{https://www.sciencedirect.com/science/article/pii/S0305054822003124}.

\bibitem[Chen et~al.(2021)Chen, Yuan, Xu, Gao, Zhang, and Liu]{CHEN2021108787}
Lei Chen, Ziyun Yuan, JianXin Xu, Jingyang Gao, Yuhan Zhang, and Gang Liu.
\newblock A novel predictive model of mixed oil length of products pipeline driven by traditional model and data.
\newblock \emph{Journal of Petroleum Science and Engineering}, 205:\penalty0 108787, 2021.
\newblock ISSN 0920-4105.
\newblock \doi{https://doi.org/10.1016/j.petrol.2021.108787}.
\newblock URL \url{https://www.sciencedirect.com/science/article/pii/S0920410521004484}.

\bibitem[Du et~al.(2022)Du, Zheng, Liang, Lu, Klemeš, Varbanov, Shahzad, Rashid, Ali, Liao, and Wang]{DU2022124689}
Jian Du, Jianqin Zheng, Yongtu Liang, Xinyi Lu, Jiří~Jaromír Klemeš, Petar~Sabev Varbanov, Khurram Shahzad, Muhammad~Imtiaz Rashid, Arshid~Mahmood Ali, Qi~Liao, and Bohong Wang.
\newblock A hybrid deep learning framework for predicting daily natural gas consumption.
\newblock \emph{Energy}, 257:\penalty0 124689, 2022.
\newblock ISSN 0360-5442.
\newblock \doi{https://doi.org/10.1016/j.energy.2022.124689}.
\newblock URL \url{https://www.sciencedirect.com/science/article/pii/S0360544222015924}.

\bibitem[Du et~al.(2023{\natexlab{a}})Du, Zheng, Liang, Liao, Wang, Sun, Zhang, Azaza, and Yan]{DU2023105647}
Jian Du, Jianqin Zheng, Yongtu Liang, Qi~Liao, Bohong Wang, Xu~Sun, Haoran Zhang, Maher Azaza, and Jinyue Yan.
\newblock A theory-guided deep-learning method for predicting power generation of multi-region photovoltaic plants.
\newblock \emph{Engineering Applications of Artificial Intelligence}, 118:\penalty0 105647, 2023{\natexlab{a}}.
\newblock ISSN 0952-1976.
\newblock \doi{https://doi.org/10.1016/j.engappai.2022.105647}.
\newblock URL \url{https://www.sciencedirect.com/science/article/pii/S0952197622006376}.

\bibitem[Du et~al.(2023{\natexlab{b}})Du, Zheng, Liang, Wang, Klemeš, Lu, Tu, Liao, Xu, and Xia]{DU2023125976}
Jian Du, Jianqin Zheng, Yongtu Liang, Bohong Wang, Jiří~Jaromír Klemeš, Xinyi Lu, Renfu Tu, Qi~Liao, Ning Xu, and Yuheng Xia.
\newblock A knowledge-enhanced graph-based temporal-spatial network for natural gas consumption prediction.
\newblock \emph{Energy}, 263:\penalty0 125976, 2023{\natexlab{b}}.
\newblock ISSN 0360-5442.
\newblock \doi{https://doi.org/10.1016/j.energy.2022.125976}.
\newblock URL \url{https://www.sciencedirect.com/science/article/pii/S0360544222028626}.

\bibitem[Du et~al.(2023{\natexlab{c}})Du, Zheng, Liang, Xia, Wang, Shao, Liao, Tu, Xu, and Xu]{DU2023128810}
Jian Du, Jianqin Zheng, Yongtu Liang, Yuheng Xia, Bohong Wang, Qi~Shao, Qi~Liao, Renfu Tu, Bin Xu, and Ning Xu.
\newblock Deeppipe: An intelligent framework for predicting mixed oil concentration in multi-product pipeline.
\newblock \emph{Energy}, 282:\penalty0 128810, 2023{\natexlab{c}}.
\newblock ISSN 0360-5442.
\newblock \doi{https://doi.org/10.1016/j.energy.2023.128810}.
\newblock URL \url{https://www.sciencedirect.com/science/article/pii/S0360544223022041}.

\bibitem[Du et~al.(2023{\natexlab{d}})Du, Zheng, Liang, Xu, Klemeš, Wang, Liao, Varbanov, Shahzad, and Ali]{DU2023127452}
Jian Du, Jianqin Zheng, Yongtu Liang, Ning Xu, Jiří~Jaromír Klemeš, Bohong Wang, Qi~Liao, Petar~Sabev Varbanov, Khurram Shahzad, and Arshid~Mahmood Ali.
\newblock Deeppipe: A two-stage physics-informed neural network for predicting mixed oil concentration distribution.
\newblock \emph{Energy}, 276:\penalty0 127452, 2023{\natexlab{d}}.
\newblock ISSN 0360-5442.
\newblock \doi{https://doi.org/10.1016/j.energy.2023.127452}.
\newblock URL \url{https://www.sciencedirect.com/science/article/pii/S0360544223008460}.

\bibitem[Du et~al.(2024)Du, Zheng, Liang, Ma, Wang, Liao, Xu, Ali, Rashid, and Shahzad]{DU2024129688}
Jian Du, Jianqin Zheng, Yongtu Liang, Yunlu Ma, Bohong Wang, Qi~Liao, Ning Xu, Arshid~Mahmood Ali, Muhammad~Imtiaz Rashid, and Khurram Shahzad.
\newblock A deep learning-based approach for predicting oil production: A case study in the united states.
\newblock \emph{Energy}, 288:\penalty0 129688, 2024.
\newblock ISSN 0360-5442.
\newblock \doi{https://doi.org/10.1016/j.energy.2023.129688}.
\newblock URL \url{https://www.sciencedirect.com/science/article/pii/S0360544223030827}.

\bibitem[He et~al.(2018)He, Lin, Wang, Liang, and Huang]{HE2018728}
Guoxi He, Mohan Lin, Baoying Wang, Yongtu Liang, and Qiyu Huang.
\newblock Experimental and numerical research on the axial and radial concentration distribution feature of miscible fluid interfacial mixing process in products pipeline for industrial applications.
\newblock \emph{International Journal of Heat and Mass Transfer}, 127:\penalty0 728--745, 2018.
\newblock ISSN 0017-9310.
\newblock \doi{https://doi.org/10.1016/j.ijheatmasstransfer.2018.08.080}.
\newblock URL \url{https://www.sciencedirect.com/science/article/pii/S001793101831250X}.

\bibitem[He et~al.(2019)He, Yang, Liao, Wang, and Sun]{https://doi.org/10.1155/2019/6892915}
Guoxi He, Na~Yang, Kexi Liao, Baoying Wang, and Liying Sun.
\newblock A novel numerical model for simulating the quantity of tailing oil in the mixed segment between two batches in product pipelines.
\newblock \emph{Mathematical Problems in Engineering}, 2019\penalty0 (1):\penalty0 6892915, 2019.
\newblock \doi{https://doi.org/10.1155/2019/6892915}.
\newblock URL \url{https://onlinelibrary.wiley.com/doi/abs/10.1155/2019/6892915}.

\bibitem[Liu et~al.(2019)Liu, Li, Cai, and Peng]{pr7010007}
Enbin Liu, Wensheng Li, Hongjun Cai, and Shanbi Peng.
\newblock Formation mechanism of trailing oil in product oil pipeline.
\newblock \emph{Processes}, 7\penalty0 (1), 2019.
\newblock ISSN 2227-9717.
\newblock \doi{10.3390/pr7010007}.
\newblock URL \url{https://www.mdpi.com/2227-9717/7/1/7}.

\bibitem[Liu et~al.(2020{\natexlab{a}})Liu, Guo, Lv, Qiao, and Azimi]{https://doi.org/10.1002/ese3.661}
Enbin Liu, Bingyan Guo, Liuxin Lv, Weibiao Qiao, and Mohammadamin Azimi.
\newblock Numerical simulation and simplified calculation method for heat exchange performance of dry air cooler in natural gas pipeline compressor station.
\newblock \emph{Energy Science \& Engineering}, 8\penalty0 (6):\penalty0 2256--2270, 2020{\natexlab{a}}.
\newblock \doi{https://doi.org/10.1002/ese3.661}.
\newblock URL \url{https://scijournals.onlinelibrary.wiley.com/doi/abs/10.1002/ese3.661}.

\bibitem[Liu et~al.(2020{\natexlab{b}})Liu, Li, Cai, Qiao, and Azimi]{doi:10.1177/0144598720911158}
Enbin Liu, Wensheng Li, Hongjun Cai, Weibiao Qiao, and Mohammadamin Azimi.
\newblock Calculation method for the amount of contaminant oil during sequential transportation through product oil pipelines.
\newblock \emph{Energy Exploration \& Exploitation}, 38\penalty0 (4):\penalty0 1014--1033, 2020{\natexlab{b}}.
\newblock \doi{10.1177/0144598720911158}.
\newblock URL \url{https://doi.org/10.1177/0144598720911158}.

\bibitem[Lu et~al.(2023{\natexlab{a}})Lu, Jiang, Xu, Ni, and Fu]{LU2023105073}
Hongfang Lu, Xinmeng Jiang, Zhao-Dong Xu, Houming Ni, and Lingdi Fu.
\newblock Mechanical behavior of high-pressure pipeline installed through horizontal directional drilling under seismic loads.
\newblock \emph{Tunnelling and Underground Space Technology}, 136:\penalty0 105073, 2023{\natexlab{a}}.
\newblock ISSN 0886-7798.
\newblock \doi{https://doi.org/10.1016/j.tust.2023.105073}.
\newblock URL \url{https://www.sciencedirect.com/science/article/pii/S0886779823000937}.

\bibitem[Lu et~al.(2023{\natexlab{b}})Lu, Jiang, Xu, Wang, and Iseley]{LU2023105077}
Hongfang Lu, Xinmeng Jiang, Zhao-Dong Xu, Niannian Wang, and David~T. Iseley.
\newblock Numerical study on mechanical properties of pipeline installed via horizontal directional drilling under static and dynamic traffic loads.
\newblock \emph{Tunnelling and Underground Space Technology}, 136:\penalty0 105077, 2023{\natexlab{b}}.
\newblock ISSN 0886-7798.
\newblock \doi{https://doi.org/10.1016/j.tust.2023.105077}.
\newblock URL \url{https://www.sciencedirect.com/science/article/pii/S0886779823000974}.

\bibitem[Lu et~al.(2023{\natexlab{c}})Lu, Xi, and Qin]{LU2023162386}
Hongfang Lu, Dongmin Xi, and Guojin Qin.
\newblock Environmental risk of oil pipeline accidents.
\newblock \emph{Science of The Total Environment}, 874:\penalty0 162386, 2023{\natexlab{c}}.
\newblock ISSN 0048-9697.
\newblock \doi{https://doi.org/10.1016/j.scitotenv.2023.162386}.
\newblock URL \url{https://www.sciencedirect.com/science/article/pii/S0048969723010021}.

\bibitem[Rachid et~al.(2002)Rachid, de~Araujo, and Baptista]{10.1115/1.1459078}
F.~B.~Freitas Rachid, J.~H.~Carneiro de~Araujo, and R.~M. Baptista.
\newblock {Predicting Mixing Volumes in Serial Transport in Pipelines }.
\newblock \emph{Journal of Fluids Engineering}, 124\penalty0 (2):\penalty0 528--534, 05 2002.
\newblock ISSN 0098-2202.
\newblock \doi{10.1115/1.1459078}.
\newblock URL \url{https://doi.org/10.1115/1.1459078}.

\bibitem[Sjenitzer(1958)]{sjenitzer1958much}
F~Sjenitzer.
\newblock How much do products mix in a pipeline?
\newblock \emph{Pipeline Engineer}, 30:\penalty0 31--34, 1958.

\bibitem[Smith et~al.(1948)Smith, Schulze, et~al.]{smith1948interfacial}
SS~Smith, RK~Schulze, et~al.
\newblock Interfacial mixing characteristics of products in products pipe line-part 1.
\newblock \emph{The Petroleum Engineer}, 20\penalty0 (8):\penalty0 330--337, 1948.

\bibitem[Wang et~al.(2020)Wang, Fan, Chin, Klemeš, and Liang]{WANG2020121831}
Bohong Wang, Yee~Van Fan, Hon~Huin Chin, Jiří~Jaromír Klemeš, and Yongtu Liang.
\newblock Emission-cost nexus optimisation and performance analysis of downstream oil supply chains.
\newblock \emph{Journal of Cleaner Production}, 266:\penalty0 121831, 2020.
\newblock ISSN 0959-6526.
\newblock \doi{https://doi.org/10.1016/j.jclepro.2020.121831}.
\newblock URL \url{https://www.sciencedirect.com/science/article/pii/S0959652620318783}.

\bibitem[Wang et~al.(2021)Wang, Klemeš, Yu, Qiu, Zheng, Lin, and Zhu]{WANG2021468}
Bohong Wang, Jiří~Jaromír Klemeš, Xiao Yu, Rui Qiu, Jianqin Zheng, Yuming Lin, and Baikang Zhu.
\newblock Planning of a flexible refined products transportation network in response to emergencies.
\newblock \emph{Journal of Pipeline Science and Engineering}, 1\penalty0 (4):\penalty0 468--475, 2021.
\newblock ISSN 2667-1433.
\newblock \doi{https://doi.org/10.1016/j.jpse.2021.12.004}.
\newblock URL \url{https://www.sciencedirect.com/science/article/pii/S2667143321000792}.
\newblock Special Issue on Smart Operation and Management of Pipelines.

\bibitem[Yu et~al.(2022)Yu, Wang, and Xu]{YU2022107613}
Li~Yu, Sujing Wang, and Qiang Xu.
\newblock Optimal scheduling for simultaneous refinery manufacturing and multi oil-product pipeline distribution.
\newblock \emph{Computers \& Chemical Engineering}, 157:\penalty0 107613, 2022.
\newblock ISSN 0098-1354.
\newblock \doi{https://doi.org/10.1016/j.compchemeng.2021.107613}.
\newblock URL \url{https://www.sciencedirect.com/science/article/pii/S0098135421003914}.

\bibitem[Yuan et~al.(2023{\natexlab{a}})Yuan, Chen, Liu, Shao, Zhang, and Ma]{YUAN2023100105}
Ziyun Yuan, Lei Chen, Gang Liu, Weiming Shao, Yuhan Zhang, and Yunxiu Ma.
\newblock Physics-informed student’s t mixture regression model applied to predict mixed oil length.
\newblock \emph{Journal of Pipeline Science and Engineering}, 3\penalty0 (1):\penalty0 100105, 2023{\natexlab{a}}.
\newblock ISSN 2667-1433.
\newblock \doi{https://doi.org/10.1016/j.jpse.2022.100105}.
\newblock URL \url{https://www.sciencedirect.com/science/article/pii/S2667143322000774}.

\bibitem[Yuan et~al.(2023{\natexlab{b}})Yuan, Chen, Liu, Shao, Zhang, and Yang]{YUAN2023211466}
Ziyun Yuan, Lei Chen, Gang Liu, Weiming Shao, Yuhan Zhang, and Wen Yang.
\newblock Physics-based bayesian linear regression model for predicting length of mixed oil.
\newblock \emph{Geoenergy Science and Engineering}, 223:\penalty0 211466, 2023{\natexlab{b}}.
\newblock ISSN 2949-8910.
\newblock \doi{https://doi.org/10.1016/j.geoen.2023.211466}.
\newblock URL \url{https://www.sciencedirect.com/science/article/pii/S2949891023000520}.

\bibitem[Zheng et~al.(2021{\natexlab{a}})Zheng, Du, Liang, Liao, Li, Zhang, and Wu]{ZHENG2021510}
Jianqin Zheng, Jian Du, Yongtu Liang, Qi~Liao, Zhengbing Li, Haoran Zhang, and Yi~Wu.
\newblock Deeppipe: A semi-supervised learning for operating condition recognition of multi-product pipelines.
\newblock \emph{Process Safety and Environmental Protection}, 150:\penalty0 510--521, 2021{\natexlab{a}}.
\newblock ISSN 0957-5820.
\newblock \doi{https://doi.org/10.1016/j.psep.2021.04.031}.
\newblock URL \url{https://www.sciencedirect.com/science/article/pii/S0957582021002172}.

\bibitem[Zheng et~al.(2021{\natexlab{b}})Zheng, Du, Liang, Wang, Liao, and Zhang]{ZHENG2021518}
Jianqin Zheng, Jian Du, Yongtu Liang, Chang Wang, Qi~Liao, and Haoran Zhang.
\newblock Deeppipe: Theory-guided lstm method for monitoring pressure after multi-product pipeline shutdown.
\newblock \emph{Process Safety and Environmental Protection}, 155:\penalty0 518--531, 2021{\natexlab{b}}.
\newblock ISSN 0957-5820.
\newblock \doi{https://doi.org/10.1016/j.psep.2021.09.046}.
\newblock URL \url{https://www.sciencedirect.com/science/article/pii/S095758202100519X}.

\bibitem[Zheng et~al.(2023)Zheng, Du, Wang, Klemeš, Liao, and Liang]{ZHENG2023113046}
Jianqin Zheng, Jian Du, Bohong Wang, Jiří~Jaromír Klemeš, Qi~Liao, and Yongtu Liang.
\newblock A hybrid framework for forecasting power generation of multiple renewable energy sources.
\newblock \emph{Renewable and Sustainable Energy Reviews}, 172:\penalty0 113046, 2023.
\newblock ISSN 1364-0321.
\newblock \doi{https://doi.org/10.1016/j.rser.2022.113046}.
\newblock URL \url{https://www.sciencedirect.com/science/article/pii/S1364032122009273}.

\end{thebibliography}

}

%%%%%%%%%%%%%%%%%%%%%%%%%%%%%%%%%%%%%%%%%%%%%%%%%%%%%%%%%%%%

\end{document}